\documentclass[a4,12pt]{article}

\usepackage{amsmath,amssymb,amsfonts,graphics,graphicx,amscd,amsfonts,epsf,epsfig}
\usepackage{setspace}
\usepackage{cite}
\usepackage{color}

\allowdisplaybreaks[4]

\date{}
\begin{document}

\begin{flushright} 
%\today\\
YITP-13-96, 
KEK-TH-1686
\end{flushright} 

\vspace{0.1cm}

\begin{center}
  {\Large
Holographic description of quantum black hole on a computer
  }
\end{center}
\vspace{0.1cm}

\vspace{0.1cm}
\begin{center}

         Masanori H{\sc anada}$^{abc}$\footnote
          {
 E-mail address : hanada@yukawa.kyoto-u.ac.jp},  
         Yoshifumi H{\sc yakutake}$^{d}$\footnote
          {
 E-mail address : hyaku@mx.ibaraki.ac.jp},  
          Goro I{\sc shiki}$^{a}$\footnote
          {
 E-mail address : ishiki@yukawa.kyoto-u.ac.jp} 
and
         Jun N{\sc ishimura}$^{ef}$\footnote
          {
 E-mail address : jnishi@post.kek.jp}

\vspace{0.5cm}

$^a${\it Yukawa Institute for Theoretical Physics, Kyoto University,\\
Kitashirakawa Oiwakecho, Sakyo-ku, Kyoto 606-8502, Japan}

$^b${\it The Hakubi Center for Advanced Research, Kyoto University,\\
Yoshida Ushinomiyacho, Sakyo-ku, Kyoto 606-8501, Japan}

$^c${\it Stanford Institute for Theoretical Physics,
Stanford University, Stanford, CA 94305, USA}

$^d${\it College of Science, Ibaraki University,
Bunkyo 1-1, Mito, Ibaraki 310-0062, Japan} 

$^e${\it KEK Theory Center, 
High Energy Accelerator Research Organization,\\
1-1 Oho, Tsukuba, Ibaraki 305-0801, Japan}

$^f${\it Graduate University for Advanced Studies (SOKENDAI),\\
1-1 Oho, Tsukuba, Ibaraki 305-0801, Japan} 

\end{center}

\vspace{1.5cm}

\begin{center}
  {\bf Abstract}
\end{center}

The discovery of the fact that black holes radiate particles
and eventually evaporate 
led Hawking to pose the well-known information loss paradox. 
This paradox caused a long and serious debate since it
claims 
that the fundamental laws of quantum mechanics may be violated.
A possible cure appeared recently from superstring theory, 
a consistent theory of quantum gravity:
if the holographic description of a quantum black hole
based on the gauge/gravity duality 
is correct, 
the information is not lost and quantum mechanics remains valid. 
Here we test this gauge/gravity duality
on a computer at the level of quantum gravity for the first time.
The black hole mass obtained
by Monte Carlo simulation of the dual gauge theory
reproduces precisely
the quantum gravity effects
in an evaporating black hole.
This result opens up totally new perspectives
towards quantum gravity
since one can simulate quantum black holes through dual gauge theories.

%%%%%%%%%%%%%%%%%%%%%%%%%%%%%%%%%%%%%%%%%%%%%%%%%%%%%%%%%%%%%%%%%%%%
%%%%%%%%%%%%%%%%%%%%%%%%%%%%%%%%%%%%%%%%%%%%%%%%%%%%%%%%%%%%%%%%%%%%
%%%%%%%%%%%%%%%%%%%%%%%%%%%%%%%%%%%%%%%%%%%%%%%%%%%%%%%%%%%%%%%%%%%%
%\section{Introduction}
%%%%%%%%%%%%%%%%%%%%%%%%%%%%%%%%%%%%%%%%%%%%%%%%%%%%%%%%%%%%%%%%%%%%
%%%%%%%%%%%%%%%%%%%%%%%%%%%%%%%%%%%%%%%%%%%%%%%%%%%%%%%%%%%%%%%%%%%%
%%%%%%%%%%%%%%%%%%%%%%%%%%%%%%%%%%%%%%%%%%%%%%%%%%%%%%%%%%%%%%%%%%%%
\newpage

\section*{Introduction}
In 1974 Hawking 
realized that a black hole should 
radiate particles as a perfect blackbody
due to quantum effects in the surrounding space,
and that the black hole should eventually
evaporate completely\cite{Hawking:1974-Nature,Hawking:1974sw}. 
This discovery
made more accurate the close analogy
between the laws of black hole physics and those of thermodynamics, 
which was pointed out originally
by Bekenstein\cite{Bekenstein:1973}.
However, it also caused a long scientific debate (see, for instance, refs~\cite{Horowitz:2003he} and~\cite{Almheiri:2012rt})  
concerning 
{\it the information loss paradox}\cite{Hawking:1976ra,Hawking:1982dj},
which can be described roughly as follows.
Suppose one throws a book
into a black hole. 
While the black hole evaporates, 
all we observe is the blackbody radiation.
Therefore, the information contained in the book
is lost forever.
This statement sharply conflicts with a basic 
consequence
of the law of quantum mechanics 
that the information of the initial state should never disappear.
Then the question is
whether the law of quantum mechanics is violated
or Hawking's argument should somehow be modified 
if full quantum effects of gravity are taken into account.

In order to resolve this paradox, 
it is necessary 
to construct microscopic states of the black hole
and to give a statistical-mechanical explanation
for the black hole entropy.
This seems quite difficult 
within general relativity 
because of the no-hair theorem, which states that 
black holes are characterized by only a few parameters.
In the mid 1990s, however, superstring theory succeeded in 
explaining the entropy of ``extremal black holes'',
a special class of black holes,
which do not evaporate\cite{Strominger:1996}.
Superstring theory contains not only strings but also
solitons called D-branes\cite{Polchinski:1995mt} as fundamental objects.
Bound states of D-branes 
can be so heavy that they look like ``black objects'' 
from a distant observer.
In fact there are many bound states,
which look like the same black hole.
These bound states can be interpreted as
the microscopic states of the black hole,
and the number of such states has been shown to 
explain precisely the black hole entropy.

\begin{figure}[htbp]
\begin{center}
\rotatebox{0}{
\scalebox{0.3}{ 
\includegraphics{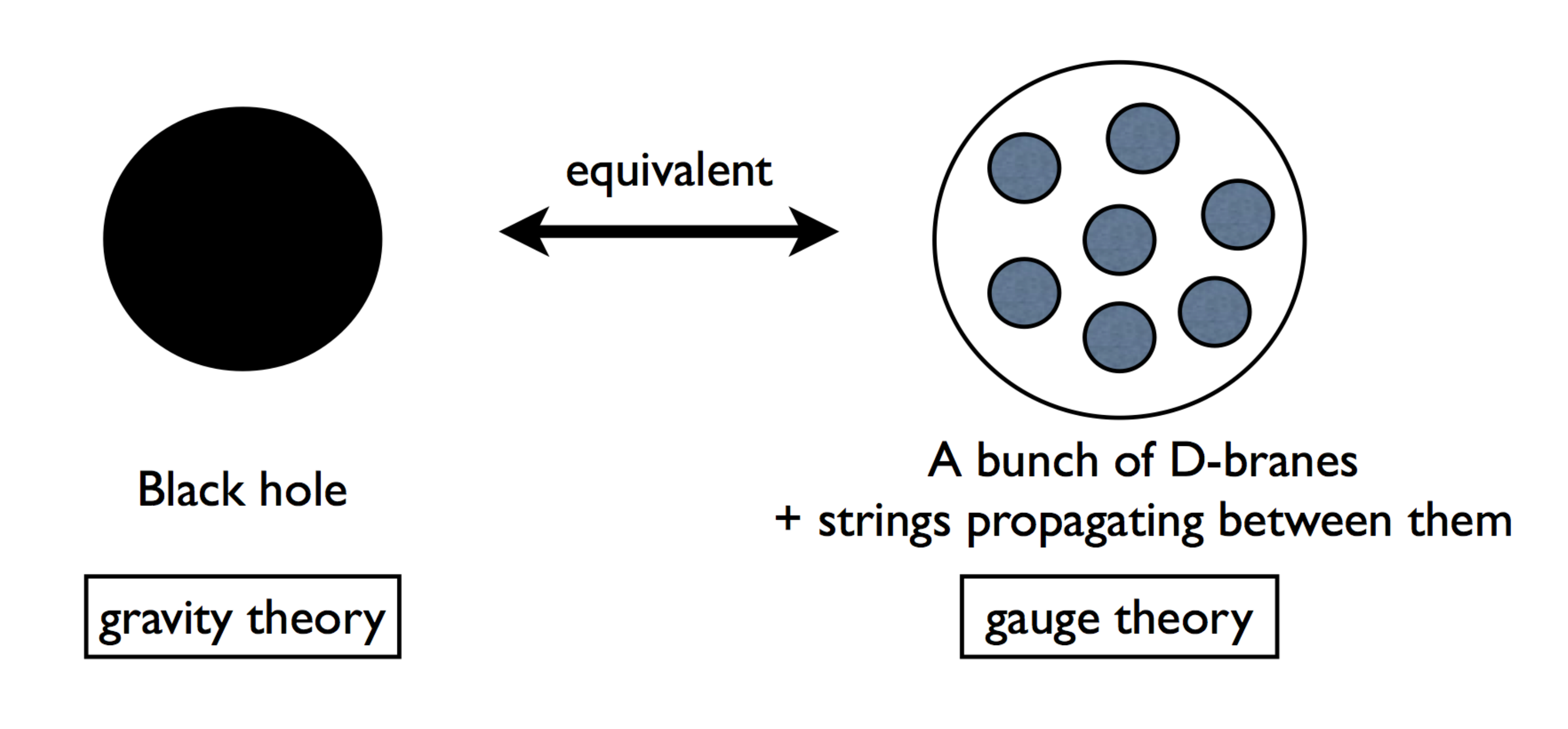}}}
\caption{ {\bf The gauge/gravity duality conjecture.}         
Black holes in superstring theory are conjectured to be described by the dual gauge theory.  }
\label{Fig1}
\end{center}
\end{figure}

However, the paradox still remains
since a complete description of an evaporating black hole
has not yet been established.
A key to really resolve the paradox is provided by
Maldacena's 
{\it gauge/gravity duality conjecture}\cite{Maldacena:1997re} (Fig.~\ref{Fig1}), 
which may be viewed as a concrete realization of 
{\it the holographic principle} 
proposed by 't Hooft\cite{'tHooft:1993gx}
and Susskind\cite{Susskind:1994vu}. 
This conjecture relates various black holes  
made of D-branes
in superstring theory to strongly coupled gauge theories,
in which the absence of information loss is manifest.
In this article 
we provide the first quantitative evidence 
for the gauge/gravity duality
at the level of quantum gravity.
We perform Monte Carlo simulation
of the dual gauge theory
in the parameter regime that corresponds to
a quantum black hole.
Our results agree precisely with
a prediction for an evaporating black hole
including quantum gravity corrections.
Thus we find that
the dual gauge theory indeed provides 
a complete description
of the quantum nature of the evaporating black hole.

%\newpage

\section*{D-particles and the gauge/gravity duality}
Superstring theory is a promising candidate 
for the theory of everything, which unifies 
the standard model of particle physics and gravity.
In particular, it provides a consistent theory of 
quantum gravity, which is otherwise difficult to formulate
due to non-renormalizable divergences at short distances.
The theory contains two kinds of strings;
closed strings and open strings.
The former mediates gravitational force, while
the latter mediates gauge interactions such as the electromagnetic force.
Superstring theory also contains solitonic objects called D-branes\cite{Polchinski:1995mt},
on which open strings can end.
The dynamical property of D-branes including the oscillation
of open strings is described by a gauge theory\cite{Witten:1995im}, 
which is a generalization of quantum electrodynamics.

Theoretical consistency requires that superstring theory should
be defined in ten-dimensional space-time.
In order to realize our four-dimensional space-time,
one can choose the size of extra six dimensions to be very small.
This procedure is called ``compactification''.
In fact there are many ways to do it without spoiling the consistency,
and by choosing the internal structure of the compactified extra dimensions 
appropriately, one can explain the variety of particles in four dimensions.  
However, since we are now interested in quantum effects of gravity,
which become important at very short distances, 
we consider superstring theory without compactification.
As a particular type of D-branes, we consider D-particles, 
which look like point-like objects in 
nine-dimensional space. 
It is known that a bunch of $N$ D-particles is described
by a gauge theory, in which all the fields are expressed
as $N\times N$ matrices depending on time\cite{Witten:1995im,Banks:1996vh,deWit:1988ig}.

Superstring theory contains only one dimensionful parameter,
which is conventionally written as
$\alpha' = \ell ^2$, where $\ell$ is the string length.
In the low-energy limit, or equivalently in the 
$\alpha' \rightarrow 0$
limit, the oscillation of closed strings is dominated
by the lowest energy states such as gravitons.
If one further neglects quantum effects,
the full superstring theory can be well approximated
by supergravity, a generalized version of Einstein's gravity theory,
which describes gravity in terms of the curvature associated with
the space-time geometry.
In supergravity,
a bunch of $N$ D-particles is expressed as
an extremal black hole, which is stable and does not cause 
Hawking radiation.
At finite temperature, the same system can 
be expressed as a non-extremal black hole.
Since it has a positive specific heat, it cools down
as it loses energy through Hawking radiation, and 
it eventually stabilizes into an extremal black hole at $T=0$.

When $N$, the number of D-particles, is large, 
the size of the black hole is large 
and the geometry is weakly curved
compared with the typical scale of quantum gravity.
Hence quantum gravity effects can indeed be neglected.
On the other hand, quantum gravity effects become important
as $N$ becomes small.
In fact such effects can make the specific heat negative.
In that case, the black hole heats up 
as it loses energy through Hawking radiation,
and it will eventually evaporate completely.
Thus this system at small $N$ is 
relevant to the information loss paradox.

Unfortunately the full quantum nature of superstring theory 
has not yet been understood.
However, according to the gauge/gravity duality conjecture,
superstring theory in the presence of the 
black hole made of D-particles is equivalent
to the gauge theory that describes the system of D-particles\cite{Itzhaki:1998dd}.
Since the gauge theory is well defined at arbitrary $N$,
it captures the full quantum nature of superstring theory
if the conjecture is correct.
Furthermore, since the gauge theory is based on 
principles of quantum mechanics, 
it is clear that the information loss does not occur
during the evaporation of the black hole.
While there are many pieces of evidence
for the gauge/gravity duality at $N=\infty$,
where classical approximation is
fully justified on the gravity side 
(See, for instance, ref~\cite{Minahan:2010js}),
very little is known about it at the level of quantum gravity.

\section*{Analysis on the gravity side}
Let us start with an analysis on the gravity side.
Readers who are not familiar with general relativity 
may jump directly to eq.~\eqref{gravity_prediction-leading}, 
which represents the outcome of this analysis.
The black hole, which is made of $N$ D-particles in superstring theory,
is described by a curved ten-dimensional space-time,
which can be obtained as a solution to the classical equation of motion
(or the ``Einstein equation'') for supergravity.
The geometry is spherically symmetric in the nine-dimensional space, 
and the black hole is surrounded by an eight-dimensional surface 
called ``event horizon''.
Once some object goes beyond the horizon from outside,
it can never come out even with the speed of light.
In particular, the metric
near the horizon is given by\cite{Horowitz:1991cd,Gibbons:1987ps}
\begin{alignat}{3}
  & ds^2 = \alpha' \Big(
- \frac{1}{\sqrt{H}} \, F dt^2 + \sqrt{H}\, \frac{1}{F} \, dU^2
  + \sqrt{H} \, U^2 d\Omega_8^2 \Big) \ , 
\label{eq:nearH}
\end{alignat}
where $U$ represents the radial coordinate and $d\Omega_8^2$
represents the line element of an eight-dimensional unit sphere.
We have introduced the functions 
$H(U) = 240\pi^5 \lambda/U^7$ and $F(U) = 1 - U_0^7/U^7$,
where the two parameters $\lambda$ and $U_0$ 
are related to the mass and charge of the black hole. 
Since $F(U)$ flips its sign at $U=U_0$, one finds that
the horizon is located at $U=U_0$.

Now we consider quantum corrections to 
the classical geometry (\ref{eq:nearH}).
Since superstring theory is defined perturbatively,
one can calculate the leading quantum corrections to the geometry,
which correspond to the $1/N^2$ corrections.
It is well-known that the scattering amplitude
involving four gravitons as asymptotic states
gives nontrivial quantum corrections to the supergravity action
at the leading order,
which include quartic terms of the Riemann tensor \cite{Gross:1986}.
By solving the equations of motion for supergravity
including such corrections, 
one obtains the metric near the horizon as (Y.~H., in preparation)
\begin{alignat}{3}
  & ds^2 
= \alpha' \Big(
 - \frac{\sqrt{H_2}}{H_1} \, 
 F_1 \, dt^2 + \sqrt{H_2} \, \frac{1}{F_1} \, dU^2
  + \sqrt{H_2} \, U^2 d\Omega_8^2 \Big)
\ , \label{eq:nearHquant}
\end{alignat}
where $H_i = H + 5\pi^{11} \lambda^3 h_i/(24U_0^{13}N^2)$ for $i=1,2$ 
and $F_1 = F + \pi^6 \lambda^2 f_1/(1152 U_0^6 N^2)$.
Here $h_i$ and $f_1$ are functions of $U/U_0$,
which can be determined uniquely.
Note that the metric (\ref{eq:nearHquant}) reduces to
(\ref{eq:nearH}) as $N\rightarrow \infty$, which corresponds to the limit
of classical gravity.
From this expression (\ref{eq:nearHquant}) for the metric, 
one finds that the position of the horizon 
is slightly shifted due to quantum effects.
One also finds that a test particle feels a repulsive force near the horizon,
which can be interpreted as the back-reaction of the Hawking radiation.

Given the geometry (\ref{eq:nearHquant}),
one can evaluate the ``energy'' $\tilde{E}$ of the black hole
as a function of temperature.
(Strictly speaking, we evaluate the difference
of the mass of the thermal non-extremal black hole
from that of the extremal one. This quantity corresponds to 
the internal energy in the dual gauge theory, hence we 
use the word ``energy''.)
For that we first calculate the entropy ${\cal S}$ of the black hole
using Wald's formula,
and obtain the ``energy'' $\tilde{E}$ by integrating 
the first law of thermodynamics, $d\tilde{E} = \tilde{T} d{\cal S}$.
Here $\tilde{T}$ denotes the Hawking temperature, which can be
derived from the geometry (\ref{eq:nearHquant}). 
Thus the ``energy'' of the black hole is evaluated as
\begin{alignat}{3}
\frac{1}{N^2} E_\text{gravity} &= 7.41 \, T^{2.8}  
  -5.77 \, T^{0.4} \frac{1}{N^2}  \ ,
\label{gravity_prediction-leading} 
\end{alignat}
up to $O(1/N^4)$ terms, where
we have introduced dimensionless parameters 
$E_\text{gravity}\equiv\lambda^{-1/3}\tilde{E}$ 
and $T\equiv\lambda^{-1/3}\tilde{T}$.
In what follows, we call the energy normalized by $\lambda^{1/3}$
``effective energy''.
The first term in (\ref{gravity_prediction-leading})
can actually be obtained\cite{Klebanov:1996un}
at the classical level from the metric (\ref{eq:nearH}), 
and the second term represents quantum gravity corrections
at the leading order.
One finds that the specific heat 
$C= dE/dT$ 
becomes negative
at sufficiently low 
$T$ 
due to the second term.
This means that the black hole becomes unstable due to the quantum
gravity effects, and it actually evaporates.

In the above analysis, we have ignored
the so-called $\alpha'$ corrections,
which represent the effects due to the oscillation of strings.
One can include these corrections to 
eq.~(\ref{gravity_prediction-leading})
as has been done in ref~\cite{Hanada:2008ez} at $N=\infty$.
Eq.~(\ref{gravity_prediction-leading}) then becomes
\begin{alignat}{3}
\frac{1}{N^2}
  E_\text{gravity}^\text{(full)} &= 
(7.41 \, T^{2.8} + a \, T^{4.6}+\cdots) +
  (-5.77 \, T^{0.4} + b \, T^{2.2}+\cdots)\frac{1}{N^2}
+ O\left(\frac{1}{N^4}\right) \ ,
\label{gravity_prediction}
\end{alignat}
where $a$ and $b$ are unknown constants.
The power of $T$ for each term 
can be determined from dimensional analysis
using some known results in superstring theory \cite{Green:2006}. 
The gauge/gravity duality claims that eq.~(\ref{gravity_prediction})
should be reproduced by the dual gauge theory\cite{Itzhaki:1998dd}.
This has been tested at $N=\infty$,
where the results from the gauge theory in the range
$0.5 \le T \le 0.7$ can indeed be nicely
fitted by the first two $O(N^0)$ terms in eq.~(\ref{gravity_prediction})
with $a=-5.58(1)$, thus
confirming the gauge/gravity duality at the level of
classical gravity\cite{Hanada:2008ez} (See also 
refs~\cite{Kabat:2000zv,Anagnostopoulos:2007fw,Catterall:2008yz}
for related works.).
The goal of our study is to see whether
the gauge theory can reproduce
the quantum gravity effects
represented by the $1/N^2$ corrections in eq.~(\ref{gravity_prediction}).

\section*{Analysis on the gauge theory side}
Let us turn to the analysis on the gauge theory side.
The gauge theory that describes 
a bunch of $N$ D-particles
is defined by the action\cite{Banks:1996vh,deWit:1988ig}
\begin{eqnarray}
S 
=
\frac{N}{\lambda} \int_0^{\beta}   
d t \, 
{\rm tr}
\bigg\{ 
\frac{1}{2} (D_t X_i)^2 - 
\frac{1}{4} [X_i , X_j]^2  
+ \frac{1}{2} \psi_\alpha D_t \psi_\alpha
- \frac{1}{2} \psi_\alpha (\gamma_i)_{\alpha\beta} 
 [X_i , \psi_\beta ]
\bigg\} \ , \label{eq:gauge}
\end{eqnarray}
where we have introduced the fields
$X_i(t)\ (i=1,2,\cdots,9)$ and 
$\psi_\alpha (t) \ (\alpha=1,2,\cdots,16)$,
which are $N\times N$ bosonic and fermionic Hermitian matrices
depending on time $t$. 
Intuitively, the diagonal elements of $X_i$ describe 
the positions of $N$ D-particles in nine spacial directions, 
and the off-diagonal elements correspond to strings 
connecting different D-particles.  
The brackets $[\ \cdot\ , \ \cdot\ ]$ represent the 
so-called commutator, 
which is defined by $[M_1,M_2]=M_1 M_2 - M_2 M_1$
for arbitrary matrices $M_1$ and $M_2$. 
We have also defined the covariant derivative 
$D_t= \partial_t -i \, [A_t,\ \cdot\ ]$,
where $A_t$ is the gauge field represented by 
an $N\times N$ Hermitian matrix.
The gamma matrices $\gamma_i\,(i=1,\cdots,9)$ are 
$16\times 16$ Hermitian matrices satisfying 
$\gamma_i\gamma_j+\gamma_j\gamma_i=2 \, \delta_{ij}$. 
As is usually done in studying thermal properties 
of gauge theories,
the time coordinate $t$ in eq.~(\ref{eq:gauge})
actually represents ``imaginary time'', which is
related to the real time $\tilde{t}$ through
$\tilde{t} = - i t$, and it is restricted to
$0 \le t \le \beta \equiv 1/\tilde{T}$, where
$\tilde{T}$ is the temperature, which should be identified
with the Hawking temperature on the gravity side.
The boundary conditions are taken to be
periodic
$A_t(t+\beta)=A_t(t),\, X_i(t+\beta)=X_i(t)$
for bosonic matrices,
and anti-periodic 
$\psi_\alpha(t+\beta)=-\psi_\alpha(t)$
for fermionic matrices.
The partition function $Z$ is defined as the sum of
the Boltzmann factors $\exp (-S)$ for all field configurations,
and the basic quantity we calculate
is the internal energy, which is defined by 
$\tilde{E} = - (\partial / \partial \beta) \log Z$.

We put the system (\ref{eq:gauge}) on a computer
as we have done in our previous 
works\cite{Anagnostopoulos:2007fw,Hanada:2008ez}.
We make a Fourier transform of each field 
with respect to time $t$,
and introduce a cutoff $\Lambda$ on the frequency.
(Strictly speaking, we need to fix the gauge symmetry 
appropriately before we introduce a cutoff.)
This method has practical advantage\cite{Hanada:2007ti}
over a more conventional method
using lattice discretization\cite{Catterall:2008yz}, in which
the matrices $X_i$ and $\psi_\alpha$ 
are put on the sites of the lattice, 
whereas the gauge fields are put on the links connecting the sites.
As far as the number of degrees of freedom is concerned,
putting the frequency cutoff $\Lambda$ corresponds to
introducing a lattice with $(2\Lambda+1)$ sites. 
In order to obtain a value in the continuum limit, 
we make an extrapolation to $\Lambda=\infty$. 
Although the fermionic matrices make the effective
Boltzmann weight complex,
we simply take the absolute value,
which is shown to be a valid approximation
in the present case\cite{Hanada:2011fq}. 

In this work we focus on small values of $N$ such as $N=3,4$ and $5$
in order to probe the quantum gravity effects, which correspond
to $1/N^2$ corrections.
This causes a new technical difficulty,
which was absent in previous works\cite{Anagnostopoulos:2007fw,Hanada:2008ez} at large $N$ 
such as $N=17$. 
We observe that the eigenvalues of the bosonic matrices $X_i$
start to diverge while we are sampling important field configurations
that contribute to the partition function.
This instability, however, can be interpreted as a physical one.
It actually corresponds to the Hawking radiation of the black hole
on the gravity side
since the black hole is microscopically described by bound states
of D-particles, and 
the positions of D-particles 
are represented by the eigenvalues of the bosonic matrices $X_i$
in the gauge theory description.
When $N$ is sufficiently large, such bound states 
are stable\cite{Anagnostopoulos:2007fw,Hanada:2008ez}, 
which reflects the stability 
of the black hole in the absence of quantum gravity effects.
When $N$ becomes small, quantum gravity effects 
destabilize the black hole.
Correspondingly, on the gauge theory side, 
we observe that the cluster of the eigenvalues becomes 
metastable as $N$ becomes small.

In order to identify the metastable bound states,
we first define a quantity
\begin{alignat}{3}
  R^2 = \frac{1}{N\beta}\int_0 ^\beta  dt
\sum_{i=1}^9{\rm tr}X_i(t)^2 \ ,
\end{alignat}
which represents the extent of the eigenvalue distribution of $X_i$.
In Fig.~2,
we show the histogram of $R^2$ for $N=4$, $T=0.10$, $\Lambda=16$.
A clear peak around $R^2 \sim 3.5$
confirms the existence of
metastable bound states, while
the non-vanishing tail at $4 \lesssim R^2 \lesssim 4.2$
reflects a run-away behavior associated with the instability.
\begin{figure}[htbp]
\begin{center}
\rotatebox{0}{\scalebox{0.5}{\includegraphics{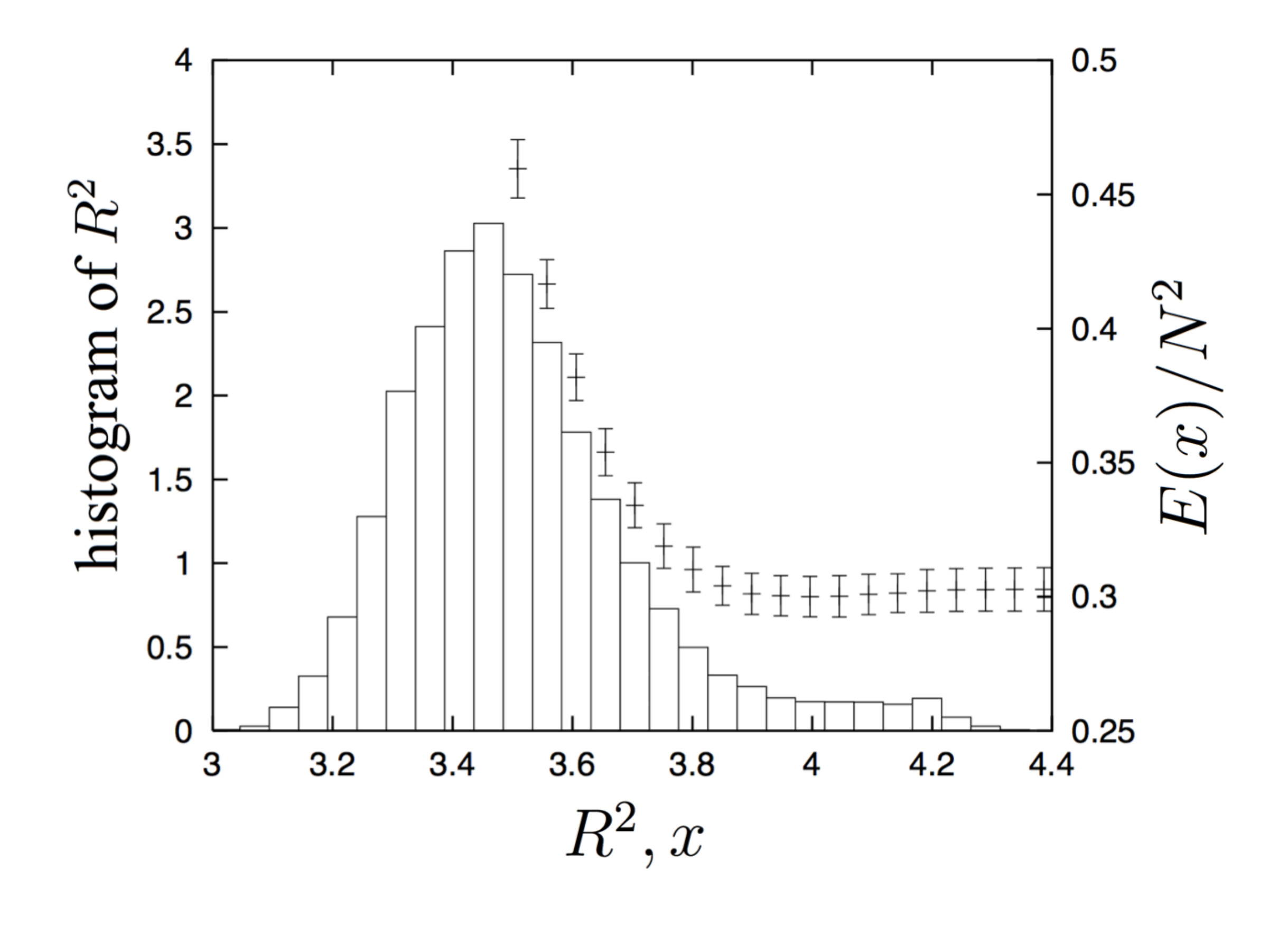}}}
\caption{ {\bf The histogram of $R^2$
and the effective internal energy $E(x)/N^2$
obtained with configurations satisfying $R^2 < x$.}
We show the results for $N=4$, $T=0.10$, $\Lambda=10$
with the choice $R^2_\text{cut}=4.2$ and $c=100$ for the cutoff
potential.
The peak of the histogram around $R^2 \sim 3.5$
represents the existence of the metastable bound states.
The plateau behavior in $E(x)/N^2$
gives us a sensible estimate of the
effective internal energy of the metastable bound states.}\label{Fig2}
\end{center}
\end{figure}

This motivates us to calculate
the effective internal energy
by using only the configurations satisfying $R^2< x $ for some $x$.
We denote such a quantity $E(x)/N^2$ and plot it also in Fig.~2.
We observe a clear ``plateau'' at the tail of
the distribution of $R^2$.
Therefore we use the height of this plateau
as a sensible estimate of the
effective internal energy of the metastable bound states.

In actual simulation we need to suppress the instability
by adding a potential term
$V_{\rm pot} = c \,  \left|R^2-R^2_\text{cut}\right|$
for $R^2>R^2_\text{cut}$
to the action (\ref{eq:gauge}),
where $c$ should be sufficiently large to kill the instability.
Note that the result for $E(x)/N^2$ presented in
Fig.~2 does not depend on $R^2_\text{cut}$ as far as
$x < R^2_\text{cut}$.
We choose $R^2_\text{cut}$
to be large enough to see the plateau behavior in $E(x)/N^2$.
For instance, Fig.~2 is obtained with
$c=100$ and $R^2_\text{cut}=4.2$.

We repeat this analysis for 
all the parameter sets $(N,T,\Lambda)$. 
We use $T=0.08$, $0.09$, $0.10$, $0.11$, $0.12$ for $N=3$, 
$T=0.07$, $0.08$, $0.09$, $0.10$, $0.11$, $0.12$ for $N=4$ and 
$T=0.08$, $0.09$, $0.10$, $0.11$ for $N=5$. 
Fitting the results $E$ obtained for finite $\Lambda$
using the ansatz $E=E_\text{gauge}+ {\rm const.}/\Lambda$, 
we obtain $E_\text{gauge}$, which represents
the effective internal energy in the continuum limit.
The fitting was made with
$\Lambda=8,10,12,14,16$ 
for $T\ge 0.10$, $\Lambda=10,12,14,16$ for $T=0.09, 0.08$  
and $\Lambda=12,14,16$ for $T=0.07$.

\begin{figure}[htbp]
\begin{center}
\rotatebox{0}{
\scalebox{0.4}
{ 
\includegraphics{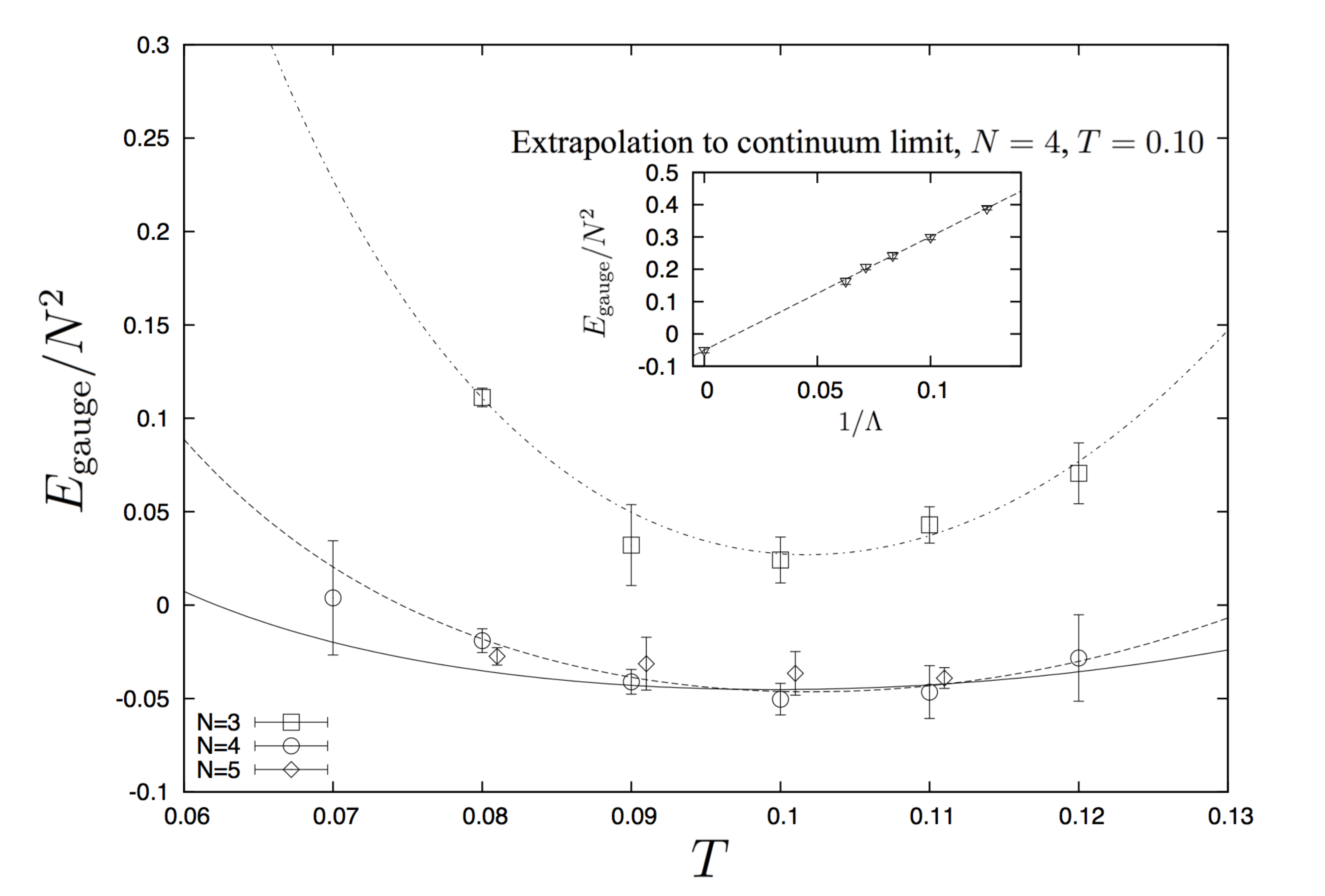}
}}
\caption{
{\bf The effective internal energy $E_\text{gauge}/N^2$ 
obtained for the metastable bound states
in the continuum limit as a function of $T$.}
Results for $N=3$ (squares), $N=4$ (circles) and $N=5$ (diamonds)
are shown.
The curves represent the fits to the behaviors 
expected from the gravity side, which shall be explained later.
The data points and the fitting curve for $N=5$ are slightly shifted 
along the horizontal axis so that the data points and the error bars 
for $N=4$ and $N=5$ do not overlap. 
In the small box, we show an extrapolation to $\Lambda=\infty$
for $N=4$ and $T=0.10$.
}
\label{Fig3}
\end{center}
\end{figure}

In Fig.~\ref{Fig3} we plot our results 
for the effective internal energy 
in the continuum limit
as a function of $T$ for $N=3,4,5$.
(In the small box we show the extrapolation to $\Lambda=\infty$
for $N=4$ and $T=0.10$ as an example.) 
The curves represent the fits to the behaviors 
expected from the gravity side, which shall be explained later.
We find that
the internal energy increases as temperature decreases,
which implies
that the specific heat is negative. 
Such a behavior is possible since
we are measuring the energy of the metastable bound states.

\section*{Testing the gauge/gravity duality}
\begin{figure}[htbp]
\begin{center}
\rotatebox{0}{
\scalebox{0.4}
{ 
\includegraphics{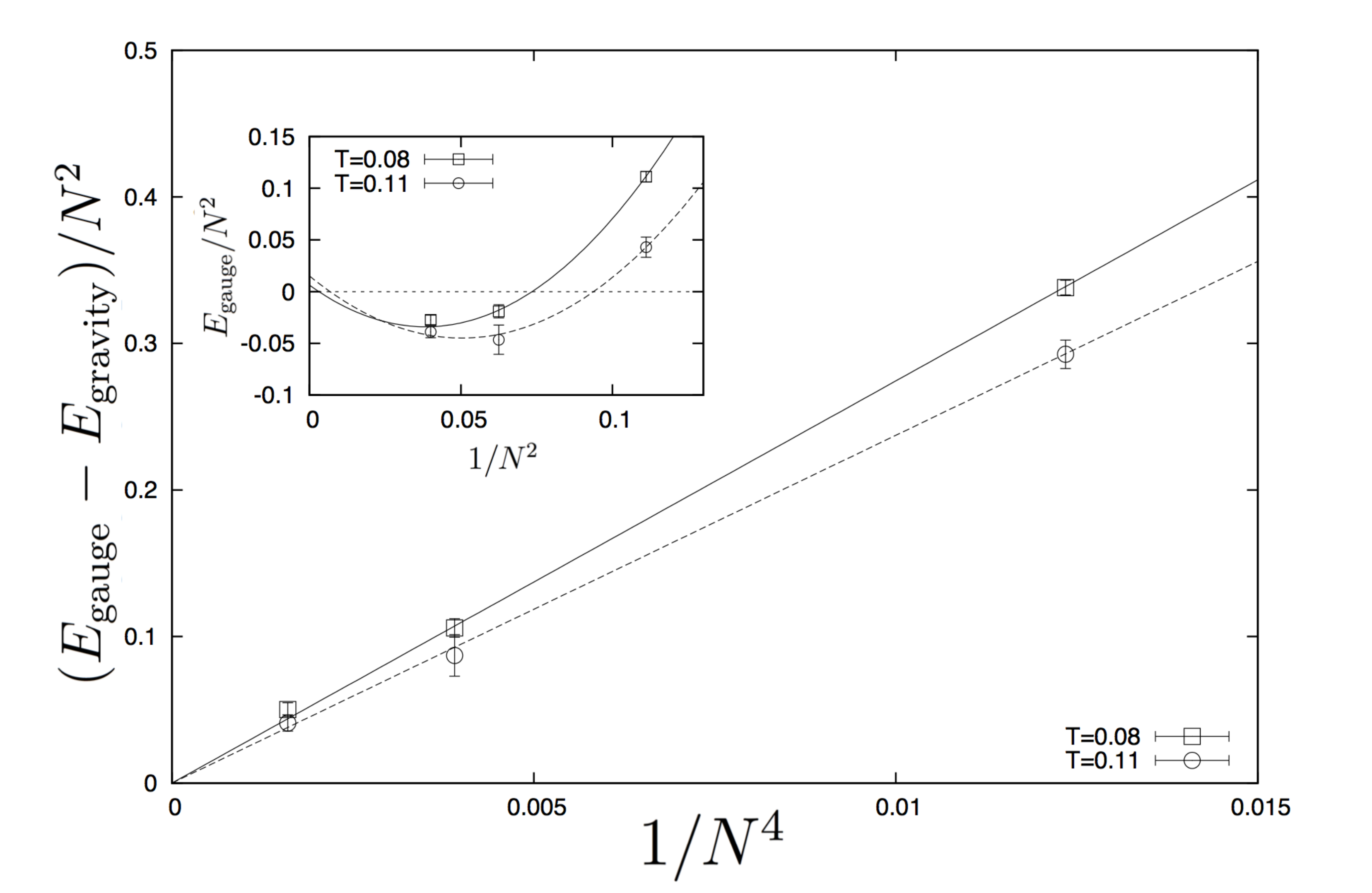}
}}
\caption{
{\bf The difference $(E_\text{gauge}-E_\text{gravity})/N^2$ as a
function of $1/N^4$.}
We show the results for $T=0.08$ (squares) and $T=0.11$ (circles). 
The data points can be nicely fitted by straight lines passing
through the origin for each $T$.
In the small box, 
we plot $E_\text{gauge}/N^2$ against $1/N^2$ for $T=0.08$ and $T=0.11$.
The curves represent the fits to the behavior
$E_\text{gauge}/N^2 =  7.41 \, T^{2.8} -5.77 \, T^{0.4}/ N^2
+{\rm const.}/N^4$
expected from the gravity side.  }
\label{Fig4}
\end{center}
\end{figure}

Now we can test the gauge/gravity duality
by comparing the results on the gauge theory side
shown in Fig.~\ref{Fig3} with 
the results on the gravity side represented by
eq.~\eqref{gravity_prediction}.
In the temperature regime $0.07\le T\le 0.12$ investigated here,
the terms with the coefficients $a$ and $b$,
which represent the $\alpha '$ corrections,
can be neglected unless $|a| \gg 700$ and $|b| \gg 500$.
(As we mentioned earlier, $a$ is obtained as $a=-5.58(1)$
by fitting the results from the gauge theory side\cite{Hanada:2008ez}
in the temperature regime $0.5 \le T \le 0.7$.)
Therefore, we can actually test
eq.~(\ref{gravity_prediction-leading}) directly.
In Fig.~\ref{Fig4}   
we plot $(E_\text{gauge}-E_\text{gravity})/N^2$
against $1/N^4$ for $T=0.08$ and $T=0.11$. 
Our data are nicely fitted by straight lines
passing through the origin.
This implies that our result obtained on the gauge theory side
is indeed consistent with the result (\ref{gravity_prediction-leading})
obtained on the gravity side including quantum gravity corrections.
In the small box of the same figure,
we plot $E_\text{gauge}/N^2$ against $1/N^2$.
The curves represent the fits to the behavior
$E_\text{gauge}/N^2 =  7.41 \, T^{2.8} -5.77 \, T^{0.4}/ N^2
+{\rm const.}/N^4$ expected from the gravity side.
We find that the $O(1/N^4)$ term is comparable
to the $O(1/N^2)$ term.
The fact that the $O(1/N^6)$ term is not visible from our data
is therefore quite nontrivial and worth being understood
from the gravity side.
The agreement of similar accuracy
is observed at other values of $T$.

\begin{figure}[htbp]
\begin{center}
\rotatebox{0}{
\scalebox{0.5}{ 
\includegraphics{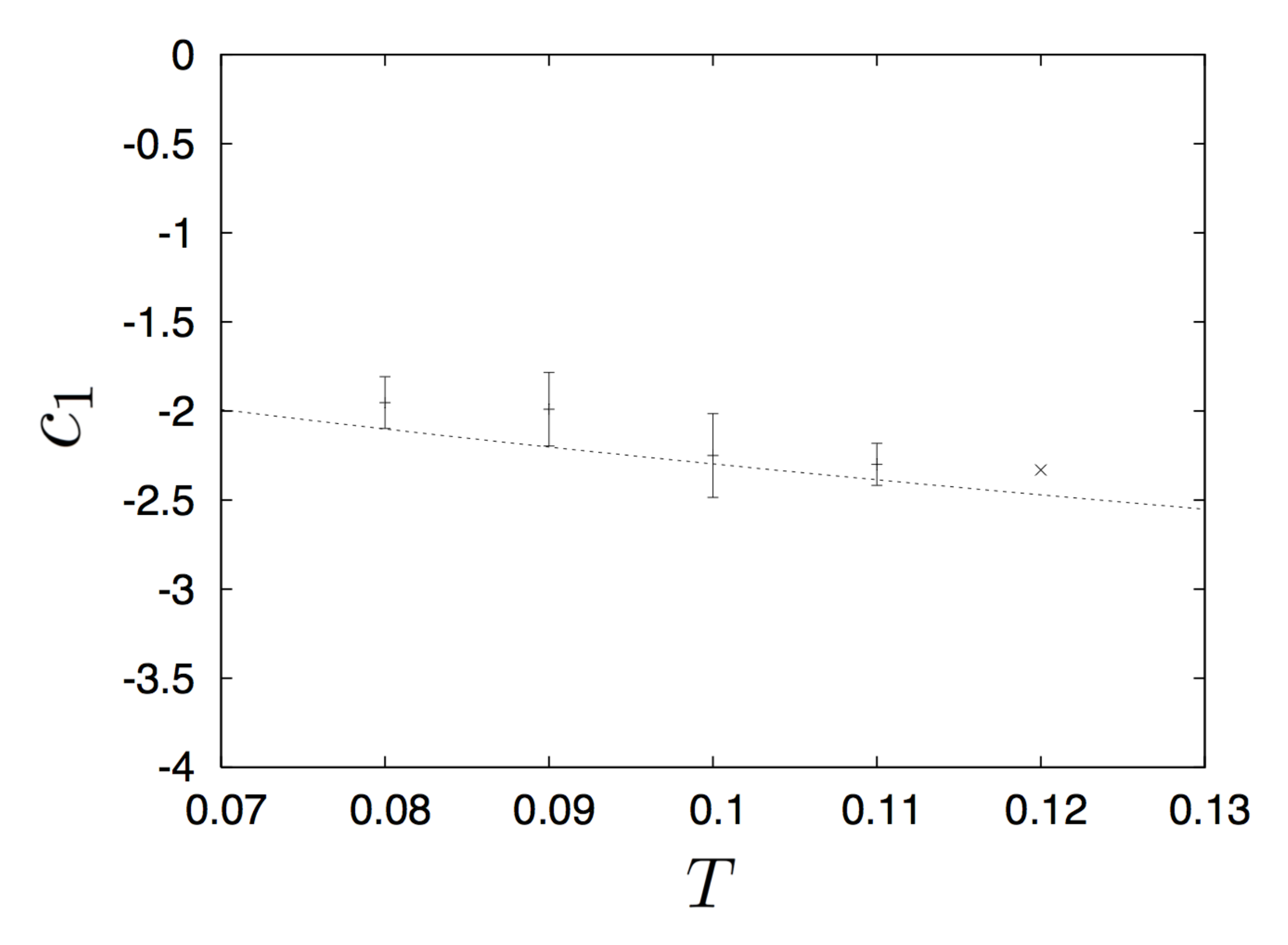}}}
\caption{{\bf The coefficient $c_1$ of the $O(1/N^2)$ term 
as a function of $T$.}  
Our results are consistent with 
the prediction $c_1 = -5.77 \, T^{0.4}$ from the gravity side (the dotted line). 
The data point at $T=0.12$ does not have an error bar 
since only two data points ($N=3,4$) were available
for making a two-parameter fit.}
\label{Fig5}
\end{center}
\end{figure}

As a further consistency check,
we have also fitted our results for each $T$
by $E_\text{gauge}/N^2 = 7.41 \, T^{2.8} + c_1/N^2+c_2/N^4$ 
leaving $c_1$ and $c_2$ as fitting parameters.
In Fig.~\ref{Fig5}, we plot $c_1$ obtained by the two-parameter
fit against $T$, which agrees well with $c_1 = -5.77 \, T^{0.4}$.

As for the coefficient $c_2$ of the $O(1/N^4)$ terms,
the prediction from the gravity side is given by
$c_2 = c \, T^{-2.6}+\cdots$, where $c$ is an unknown constant.
In fact $c_2$ can be fitted, 
for instance, by $c_2 = c \, T^{-2.6}+ \tilde{c} \, T^{p}$
with $c=0.0340(12)$, $\tilde{c}=0.17(23)\times 10^6$ and $p=4.30(62)$. 
(The value for $\tilde{c}$ looks huge, but it is
actually compensated by the high power of $T$ within
the temperature region investigated here.)
Therefore we consider that the $T$ dependence of $c_2$ is 
also consistent with the prediction from the gravity side.
The curves in Fig.~\ref{Fig3} represent
$E_\text{gauge}/N^2= 
E_\text{gravity}/N^2 +(c \, T^{-2.6}+ \tilde{c} \, T^{p})/N^4$
with the fitting parameters obtained above.

\section*{Summary and discussions}
In this article we have given quantitative evidence 
for the gauge/gravity duality at the level of quantum gravity. 
In particular, we find that 
an evaporating black hole can be described by
the dual gauge theory, which is based on
fundamental principles of quantum mechanics.
This provides us with an explicit example in which
the information is not lost in an evaporating black hole.

Our work suggests a new approach
to the quantum nature of gravity.
Since the gauge/gravity duality
is confirmed including quantum gravity effects,
we can study various issues involving quantum gravity
by using Monte Carlo simulation of the dual gauge theory.
Thus the situation has become quite close to
the studies of the strong interaction by simulating
Quantum Chromodynamics on the lattice,
which successfully explained 
the mass spectrum of hadrons\cite{Durr:2008zz} and 
the nuclear force\cite{Ishii:2006ec} recently.
We can now apply essentially the same method
to study quantum gravity.

\section*{Acknowledgements}  
The authors would like to thank Sinya Aoki, Sean Hartnoll, 
Issaku Kanamori, Hikaru Kawai, Erich Poppitz, 
Andreas Sch\"{a}fer, Stephen Shenker,
Leonard Susskind, Masaki Tezuka, Akiko Ueda and Mithat \"{U}nsal for 
discussions and comments.
M.~H.\ is supported by the Hakubi Center 
for Advanced Research, Kyoto University
and by the National Science Foundation 
under Grant No.\ PHYS-1066293.
M.~H.\ and Y.~H.\ are partially supported 
by the Ministry of Education, Science, 
Sports and Culture, Grant-in-Aid for Young Scientists (B), 
25800163, 2013 (M.~H.), 19740141, 2007 (Y.~H.) and 24740140, 2012 (Y.~H.).
The work of J.~N.\ was
supported in part by Grant-in-Aid for Scientific Research
(No.\ 20540286, 23244057)
from Japan Society for the Promotion of Science.
Computations were carried out
on PC cluster systems in KEK and Osaka University Cybermedia Center 
(the latter being provided by 
the HPCI System Research Project, project ID:hp120162).

%%%%%%%%%%%%%%%%%%%%%%%%%%%%%%%%%%%%%%%%%%%%%%%%%%%%%%%%%%%%%%%%%%%%

\end{document}